\documentclass[reqno]{amsart} % 12pt, {article} %
\usepackage{epsfig}
\usepackage{graphicx}% Include figure files
\usepackage{bm}% bold math
\usepackage{amssymb}

\newcommand{\nc}{\newcommand}
\newcommand{\Z}{\mathbb Z}
\nc{\rnc}{\renewcommand} \nc{\beq}{\begin{equation}}
\nc{\eeq}{\end{equation}} \nc{\beqa}{\begin{eqnarray}}
\nc{\eeqa}{\end{eqnarray}}
\def \z{\underline{z}}

\def \s{\underline{s}}
\def \y{\underline{y}}
\def \w{\underline{w}}

\def \x{\underline{x}}
\def \z{\underline{z}}
\def \y{\underline{y}}

\def \t{\underline{t}}
\def \s{\underline{s}}
\def \T{\theta}

\def \J{b}
\def \K{c}

\newtheorem{theorem}{Theorem}

\begin{document}

\begin{flushright} DESY 12-048 \end{flushright}

\title[ 't Hooft anomaly matching conditions]
{\bf Elliptic hypergeometric integrals and
\\ 't Hooft anomaly matching conditions}

\author{V.~P.~Spiridonov}
\address{Bogoliubov Laboratory of Theoretical Physics,
JINR, Dubna, Moscow Region 141980, Russia; e-mail
address: spiridon@theor.jinr.ru}

\author{G.~S.~Vartanov}
\address{DESY Theory, Notkestrasse 85, 22603 Hamburg, Germany;
e-mail address: \linebreak grigory.vartanov@desy.de}

\begin{abstract}
Elliptic hypergeometric integrals describe superconformal indices
of $4d$ supersymmetric field theories.
We show that all 't Hooft anomaly matching conditions for Seiberg dual
theories can be derived from $SL(3,\mathbb{Z})$-modular transformation
properties of the kernels of dual indices.
\end{abstract}

\maketitle

%\tableofcontents

\begin{flushright} \textit{To the memory of F. A. Dolan} \end{flushright}

\section{Superconformal index}

In a remarkable paper \cite{Dolan:2008qi} Dolan and Osborn
have recognized the fact that superconformal indices (SCIs)
of $4d$ supersymmetric gauge theories
\cite{Kinney,Romelsberger1,Romelsberger2}
are expressed in terms of elliptic hypergeometric integrals (EHIs)
\cite{S1,S2} (see \cite{S3} for a review). This observation provides currently
the most rigorous mathematical confirmation of $\mathcal{N}=1$  Seiberg
electric-magnetic duality \cite{Seiberg} through the equality of
dual indices.
In a sequel of papers \cite{SV1,SV2,SV3,SV4,V,S5,SV5} we have
systematically studied
this interrelation between SCIs and EHIs and described many
new $\mathcal{N}=1$
physical dualities and conjectured many new identities for EHIs.
Supersymmetric field theories on curved backgrounds
and corresponding indices modeling SCIs have been studied in \cite{Sen,FS,Demail}.
The theory of EHIs was applied also to a description of the $S$-duality conjecture
for $\mathcal{N}=2,4$ extended supersymmetric field theories
\cite{GPRR1,GPRR2,SV2,SV4,Gadde:2011ik,Gadde:2011uv}.
Several modifications of SCIs have been considered recently such as the inclusion
of charge conjugation \cite{Zwiebel:2011wa}, indices on lens spaces
\cite{Benini:2011nc}, inclusion of surface operators \cite{Nakayama:2011pa}
or line operators \cite{Dimofte:2011py,Gang:2012yr}. In \cite{S5} it was shown
that SCIs of $4d$ theories describe partition functions of some
novel integrable models of $2d$ spins systems where the Seiberg duality plays
the role of Kramers-Wannier duality transformations.

By definition SCIs count BPS states
protected by one supersymmetry which can not be combined to form
long multiplets. The $\mathcal{N}=1$ superconformal algebra
of $SU(2,2|1)$ space-time symmetry group is generated by $J_i,
\overline{J}_i$ (Lorentz rotations), $P_\mu$ (translations),
$K_\mu$, (special conformal transformations),
$H$ (dilatations) and $R$ ($U(1)_R$-rotations). In addition
to the bosonic generators there are supercharges
$Q_{\alpha},\overline{Q}_{\dot\alpha}$ and their superconformal
partners $S_{\alpha},\overline{S}_{\dot\alpha}$.
Distinguishing a pair of supercharges, say,
$Q=\overline{Q}_{1 }$ and $Q^{\dag}=-{\overline S}_{1}$, one has
\begin{equation}
\{Q,Q^{\dag}\}= 2{\mathcal H},\quad Q^2=\left(Q^\dag\right)^2=0,\qquad
\mathcal{H}=H-2\overline{J}_3-3R/2. \label{susy}\end{equation}
The superconformal index is defined now by the trace
\begin{eqnarray}  && \makebox[-2em]{}
I(p,q,f_k) =  \text{Tr} \Big( (-1)^{\mathcal F}
p^{\mathcal{R}/2+J_3}q^{\mathcal{R}/2-J_3}
 e^{\sum_{k}f_kF_k}e^{-\beta {\mathcal H}}\Big),
\quad \mathcal{R}= R+2\overline{J}_3,
\label{Ind}\end{eqnarray}
where ${\mathcal F}$ is the fermion number operator.
Chemical potentials $f_k$ are the group
parameters of the flavor symmetry group with the maximal torus
generators $F_k$. The variables $p$ and $q$ are fugacities
(group parameters) for the operators $\mathcal{R}/2\pm J_3$
commuting with $Q$ and $Q^{\dag}$. Only
zero modes of $\mathcal H$ contribute to the trace because
relation (\ref{susy}) is preserved by the operators used in
(\ref{Ind}).

An explicit computation of SCIs for ${\mathcal
N}=1$ theories results in the prescription
\cite{Romelsberger2,Dolan:2008qi} according to which one should first
compute the trace in (\ref{Ind}) over the single particle states
\begin{eqnarray}\label{index}
\text{ind}(p,q,\z,\y) &=& \frac{2pq - p - q}{(1-p)(1-q)}
\chi_{adj}(\z)\cr &+& \sum_j
\frac{(pq)^{R_j/2}\chi_{R_F,j}(\y)\chi_{R_G,j}(\z) -
(pq)^{1-R_j/2}\chi_{{\bar R}_F,j}(\y)\chi_{{\bar R}_G,j}(\z)}{(1-p)(1-q)},
\nonumber\end{eqnarray}
where the first term describes the
contribution of gauge superfields lying in the adjoint
representation of the gauge group $G_c$. The sum over $j$ corresponds to
the contribution of chiral matter superfields $\Phi_j$ transforming as the gauge
group representations $R_{G,j}$ and flavor symmetry
group representations $R_{F,j}$. The functions $\chi_{adj}(\z)$,
$\chi_{R_F,j}(\y)$ and $\chi_{R_G,j}(\z)$ are the corresponding
characters.  The exponents $R_j$ are the field $R$-charges.
To obtain the full SCI, this single
particle states index is inserted into the ``plethystic" exponential
with the subsequent averaging over the gauge group leading to the matrix integral
\begin{equation}\label{Ind_fin}
I(p,q,\y) \ = \ \int_{G_c} d \mu(\z)\, \exp \bigg ( \sum_{n=1}^{\infty}
\frac 1n \text{ind}\big(p^n ,q^n, \z^n , \y^ n\big ) \bigg ),
\end{equation}
where $d \mu(\z)$ is the $G_c$-invariant measure.

Let us take the initial Seiberg duality for SQCD \cite{Seiberg}
and consider it in detail. Namely, take a $4d$ $\mathcal{N}=1$
SYM theory with $G_c=SU(N_c)$
gauge group and $SU(N_f)_l \times SU(N_f)_r \times U(1)_B$
flavor symmetry group. The original (electric) theory has $N_f$ left
and $N_f$ right quarks $Q$ and $\widetilde{Q}$ lying in the
fundamental and antifundamental representation of $SU(N_c)$ and
having $+1$ and $-1$ baryonic charges and the $R$-charge
$R=(N_f-N_c)/N_f$ (this is the $R$-charge for the
scalar component, the $R$-charge of the fermion component
is $R-1$). The field content of the described theory is
summarized in the following table

\begin{center}
\begin{tabular}{|c|c|c|c|c|c|}
\hline
& $SU(N_c)$ & $SU(N_f)_l$ & $SU(N_f)_r$ & $U(1)_B$ & $U(1)_R$ \\
\hline
$Q$ & $f$ & $f$ & 1 & 1 & $\frac{N_f-N_c}{N_f}$ \\
$\widetilde{Q}$ & $\overline{f}$ & 1 & $\overline{f}$
                               & $-1$ & $\frac{N_f-N_c}{N_f}$ \\
$V$ & $adj$ & 1 & 1 & 0 & 1 \\
\hline
\end{tabular}
\end{center}

Corresponding SCI is given by the following elliptic hypergeometric
integral \cite{Dolan:2008qi}
\begin{eqnarray}\label{IE-seiberg}
&& I_E =   \kappa_{N_c} \int_{\mathbb{T}^{N_c-1}}
 \frac{\prod_{i=1}^{N_f} \prod_{j=1}^{N_c}
    \Gamma(s_i z_j,t^{-1}_i z^{-1}_j;p,q)}
{\prod_{1 \leq i < j \leq N_c} \Gamma(z_i z^{-1}_j,z_i^{-1}
z_j;p,q)} \prod_{j=1}^{N_c-1} \frac{d z_j}{2 \pi \textup{i} z_j},
\end{eqnarray}
where $\mathbb{T}$ denotes the unit circle with positive orientation,
$\prod_{j=1}^{N_c} z_j =1$, $|s_i|, |t_i^{-1}|<1$, and the balancing condition reads
$ST^{-1} = (pq)^{N_f-N_c}$ with
$S =\prod_{i=1}^{N_f}s_i,$ $T=\prod_{i=1}^{N_f}t_i.$
Here we introduced the parameters $s_i$ and $t_i$ as
\begin{equation}
s_i=(pq)^{R/2}vx_i, \qquad t_i=(pq)^{-R/2}vy_i,
\label{ini_var}\end{equation}
where $v$, $x_i$ and $y_i$ are fugacities for $U(1)_B$,
$SU(N_f)_l$ and $SU(N_f)_r$ groups, respectively, with the constraints
$\prod_{i=1}^{N_f}x_i=\prod_{i=1}^{N_f}y_i=1$, and
$$
\kappa_{N_c} = \frac{(p;p)_{\infty}^{N_c-1} (q;q)_{\infty}^{N_c-1}}{N_c!},
\qquad
(a;q)_\infty=\prod_{k=0}^\infty(1-aq^k).
$$
We use also conventions
$$
\Gamma(a,b;p,q):=\Gamma(a;p,q)\Gamma(b;p,q),\quad
\Gamma(az^{\pm1};p,q):=\Gamma(az;p,q)\Gamma(az^{-1};p,q),
$$
where
\beq \label{ellg}
\Gamma(z;p,q)= \prod_{i,j=0}^\infty
\frac{1-z^{-1}p^{i+1}q^{j+1}}{1-zp^iq^j}, \quad |p|, |q|<1,
\eeq
is the (standard) elliptic gamma function.

According to \cite{Seiberg} the dual (magnetic) theory  is described by
a $4d$ $\mathcal{N}=1$ SYM theory with the gauge group
$\widetilde{G}_c=SU(\widetilde{N}_c)$, $\widetilde{N}_c=N_f-N_c,$ sharing the same
flavor symmetry. Here one has dual quarks $q$ and $\widetilde{q}$
lying in the fundamental and antifundamental
representation of $\widetilde{G}_c$ with $U(1)_B$-charges
$N_c/(N_f-N_c)$ and $-N_c/(N_f-N_c)$ and the $R$-charges $N_c/N_f$.
Additionally, there are mesons -- singlets of $\widetilde{G}_c$ lying in the
fundamental representation of $SU(N_f)_l$ and antifundamental
representation of $SU(N_f)_r$ ($M_i^j=Q_i\widetilde{Q}^j,
i,j=1,\ldots,N_f$). It is convenient to collect again all fields data
in one table
\begin{center}
\begin{tabular}{|c|c|c|c|c|c|}
\hline
& $SU(\widetilde{N}_c)$ & $SU(N_f)_l$ & $SU(N_f)_r$ & $U(1)_B$ & $U(1)_R$ \\
\hline
$M$ & 1 & $f$ & $\overline{f}$ & 0 & $2\frac{\widetilde{N}_c}{N_f}$ \\
$q$ & $f$ & $\overline{f}$ & 1 & $\frac{N_c}{\widetilde{N}_c}$ & $\frac{N_c}{N_f}$ \\
$\widetilde{q}$ & $\overline{f}$ & 1 & $f$ & $-\frac{N_c}{\widetilde{N}_c}$ & $\frac{N_c}{N_f}$ \\
$\widetilde V$ & $adj$ & 1 & 1 & 0 & 1 \\
\hline
\end{tabular}
\end{center}

These two SQCD-type theories are dual to each other in their infrared fixed points
when the magnetic theory has a dynamically generated superpotential
\cite{Seiberg}, $W_{dyn} = M_i^j q^i \widetilde{q}_j$. The SCI
of the magnetic theory is
\begin{eqnarray}\label{IM-seiberg}
&&    I_M = \kappa_{N_{\widetilde{N}_c}} \prod_{1 \leq
i,j \leq N_f}
    \Gamma(s_i t^{-1}_j;p,q)
 \\ \nonumber && \makebox[3em]{} \times \int_{\mathbb{T}^{\widetilde{N}_c-1}}
     \frac{\prod_{i=1}^{N_f} \prod_{j=1}^{\widetilde{N}_c}
     \Gamma(S^{1/\widetilde{N}_c} s_i^{-1} \widetilde{z}_j,
  T^{-1/\widetilde{N}_c} t_i \widetilde{z}_j^{-1};p,q)}
{\prod_{1 \leq i < j \leq \widetilde{N}_c}
     \Gamma(\widetilde{z}_i \widetilde{z}_j^{-1},\widetilde{z}_i^{-1}
\widetilde{z}_j;p,q)}\prod_{j=1}^{\widetilde{N}_c-1}  \frac{d \widetilde{z}_j}
{2 \pi \textup{i} \widetilde{z}_j},
\end{eqnarray}
where $\prod_{j=1}^{\widetilde{N}_c} \widetilde{z}_j = 1$,
and  $|S^{1/\widetilde{N}_c} s_i^{-1}|,|T^{-1/\widetilde{N}_c} t_i |<1$.
As discovered by Dolan and Osborn  \cite{Dolan:2008qi},
the equality of SCIs $I_E=I_M$ coincides with the mathematical
identity established for $N_c=2, N_f=3, 4$ in \cite{S1,S2,S3} and for
arbitrary parameters in \cite{Rains}.

In \cite{SV2,SV3} we proposed to relate known physical checks of the
Seiberg duality to certain mathematical properties of EHIs:
\begin{enumerate}
  \item 't Hooft anomaly matching conditions for dual theories
\cite{Hooft} were conjectured to follow from
the so-called total ellipticity property for elliptic
hypergeometric terms \cite{S3,spi:cirm}.
  \item
One can reduce the duality from $N_f$ to $N_f-1$ flavors
by integrating out one flavor. At the level of SCIs this can be realized by
the restriction $s_ft_f^{-1}=pq$ for fugacities of the flavor $f$
one wants to integrate out.
  \item Matching of the moduli spaces and gauge invariant operators
should correspond to the equality of coefficients in the series expansions
of SCIs having a topological meaning.
\end{enumerate}

The main purpose of this paper is to analyze in detail the
first point of this list. Namely, we show that our original
conjecture (1) is false, i.e. the total ellipticity condition is
not sufficient to match all anomalies. Instead, \textit{all}
continuous current anomalies match as a consequence of
the nontrivial $SL(3,\mathbb{Z})$-modular group properties
of the kernels of elliptic hypergeometric integrals describing indices.
The importance of this modular group for dualities was announced by
us in \cite{DSV}.

\section{The modified elliptic gamma function}

We start from describing the modified elliptic gamma function
\cite{S2} playing a key role in our considerations.
Function \eqref{ellg} satisfies the following equations
\begin{equation}
\Gamma(qz;p,q) = \theta(z;p) \Gamma(z;p,q), \ \ \ \ \
\Gamma(pz;p,q) = \theta(z;q) \Gamma(z;p,q),
\label{gam_eq}\end{equation}
where $\theta(z;p)$ is a theta-function
$$
\theta(z;p) = (z;p)_\infty (pz^{-1};p)_\infty.
$$
This (shortened) theta function satisfies symmetry properties
$$
\theta(pz;p)=\theta(z^{-1};p)=-z\theta(z;p),
$$
and for any $k\in\Z$
\begin{equation}
\theta(p^kz;p) = (-z)^{-k} p^{-k(k-1)/2} \theta(z;p).
\label{quasi}\end{equation}

Equations \eqref{gam_eq} necessarily demand that $|p|, |q|<1$,
and for $p^n\neq q^m,\, n,m\in\mathbb{Z}$, they define
$\Gamma(z;p,q)$ uniquely as a meromorphic function
of $z\in\mathbb{C}^*$ with the normalization $\Gamma(\sqrt{pq};p,q)=1$.

Let us take three complex variables
$\omega_1, \omega_2, \omega_3$ and define the bases
\beq \label{def_basis}
p \ = \ e^{2 \pi \textup{i}
\omega_3/\omega_2}, \quad q \ = \ e^{2 \pi \textup{i}
\omega_1/\omega_2}, \quad r \ = \ e^{2 \pi \textup{i} \omega_3/\omega_1}
\eeq
together with their particular modular transformed partners
\beq \widetilde{p} \ = \ e^{-2 \pi \textup{i}
\omega_2/\omega_3}, \quad \widetilde{q} \ = \ e^{-2 \pi \textup{i}
\omega_2/\omega_1}, \quad \widetilde{r} \ = \ e^{-2 \pi \textup{i}
\omega_1/\omega_3}. \eeq

Remind now that the elliptic gamma function is originally related
to one finite difference equation
\beq \label{eq1}
f(u+\omega_1) = \theta(e^{2 \pi \textup{i} u/\omega_2};p) f(u), \ u \in \mathbb{C}.
\eeq
It coincides with the first equation above for $\Gamma(z;p,q)$ with
$z=e^{2 \pi \textup{i} u/\omega_2}$, but it assumes only one
constraint for bases, $|p|<1$. There exist other nontrivial solutions
to (\ref{eq1}) which do not require $|q|<1$. Namely, consider
equation (\ref{eq1}) together with two additional equations
\beq\label{eq23}
f(u+\omega_2) = \theta(e^{2 \pi \textup{i} u/\omega_1};r) f(u), \ \ \ \ \
f(u+\omega_3) =e^{-\pi\textup{i}B_{2,2}(u;\mathbf{\omega})} f(u),
\eeq
where $B_{2,2}(u;\mathbf{\omega})$ is a second order Bernoulli polynomial
$$
B_{2,2}(u;\mathbf{\omega})=\frac{u^2}{\omega_1\omega_2}
-\frac{u}{\omega_1}-\frac{u}{\omega_2}+
\frac{\omega_1}{6\omega_2}+\frac{\omega_2}{6\omega_1}+\frac{1}{2}.
$$
Then for incommensurate $\omega_j$'s the modified elliptic gamma function \cite{S2}
\beq \label{MEGF1}
\mathcal{G}(u;{\bf \omega}) \ = \ \Gamma(e^{2 \pi \textup{i}
u/\omega_2};p,q) \Gamma(r e^{-2 \pi \textup{i}
u/\omega_1};\widetilde{q},r)
\eeq
defines the unique meromorphic solution of these three equations
satisfying the normalization condition
$\mathcal{G}(\sum_{i=1}^3\omega_i/2;{\bf \omega})=1$.
This is a meromorphic function of $u$ even for $\omega_1/\omega_2>0$,
when $|q|=1$, which is easily seen from its another representation
 \beq \label{MEGF2}
\mathcal{G}(u;{\bf \omega}) \ = \ e^{-\frac{\pi
\textup{i}}{3} B_{3,3}(u;\mathbb{\omega})} \Gamma(e^{-2 \pi \textup{i}
u/w_3};\widetilde{r},\widetilde{p}), \eeq
where $B_{3,3}$ is a Bernoulli polynomial of the third order
\beqa \label{B33} && \makebox[-2em]{}
B_{3,3}(u;\mathbb{\omega}) = \frac{1}{\omega_1\omega_2 \omega_3}
\Bigl(u-\frac12\sum_{k=1}^3\omega_k\Bigr)\Bigl((u-\frac12\sum_{k=1}^3\omega_k )^2
-\frac14 \sum_{k=1}^3\omega_k^2\Bigr).
\eeqa
Multiple Bernoulli polynomials are defined in the theory of Barnes
multiple zeta-function \cite{S3} from the following expansion
$$
\frac{x^m e^{xu}}{\prod_{k=1}^m(e^{\omega_k x}-1)}
=\sum_{n=0}^\infty B_{m,n}(u;\omega_1,\ldots,\omega_m)\frac{x^n}{n!}.
$$

The equality of expressions \eqref{MEGF1} and \eqref{MEGF2} follows
from the Jacobi theorem on absence of nontrivial triply periodic functions,
since both expressions satisfy three equations and the normalization
condition indicated above. This equality represents a modular transformation
law from the $SL(3,\mathbb{Z})$-group \cite{FV}.
We stress that all three bases $p,q,r$ are involved into
the description of $\mathcal{G}(u;{\bf \omega})$. The modified
elliptic hypergeometric integrals built from the modified elliptic
gamma functions \cite{unit} and the Bernoulli polynomial (\ref{B33})
will play the major role in our analysis below.

We shall use also the well known modular transformation properties
of the theta function
\begin{equation}
\theta(e^{-2\pi\textup{i} u/\omega_1};e^{-2\pi\textup{i} \omega_2/\omega_1)}
=e^{\pi \textup{i}B_{2,2}(u;\omega_1,\omega_2)}\theta(e^{2\pi\textup{i} u/\omega_2};
e^{2\pi\textup{i}  \omega_1 /\omega_2})
\label{mod_theta}\end{equation}
and of the Dedekind eta-function
\beq \label{Ded}
e^{-\frac{\pi \textup{i}}{12 \tau}} (e^{-2 \pi \textup{i}/\tau};
e^{-2 \pi \textup{i}/\tau})_\infty=
(-\textup{i} \tau)^{1/2}
e^{\frac{\pi \textup{i} \tau}{12}} (e^{2 \pi \textup{i} \tau};
e^{2 \pi \textup{i} \tau})_\infty,
\eeq
where $\sqrt{-\textup{i}}=e^{-\pi\textup{i}/4}$.

\section{The total ellipticity condition and anomaly matchings}

\begin{figure}[ht]\vspace{0.3cm}
\begin{center}
\leavevmode \epsfxsize=5cm \epsffile{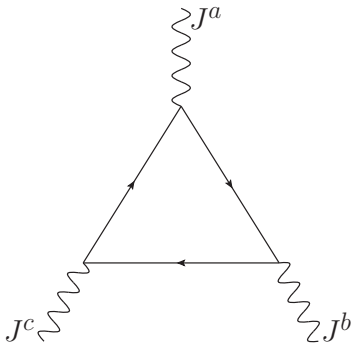}
\end{center}\vspace{-0.2cm}
\caption{A one-loop triangle diagram describing anomalies.}\label{AnomD}
\end{figure}

{\bf Anomalies.}
We would like to remind basic facts about the anomalies and 't Hooft
anomaly matching conditions. All continuous symmetry anomalies are obtained from
the one-loop triangular diagrams presented in Fig. \ref{AnomD} where
loop lines contain all possible fermions and external lines are either
global symmetry currents, gauge currents or graviton currents
(there are two such diagrams: the second one is obtained
from Fig. \ref{AnomD} by reversing the fermion current). When all
external lines describe gauge currents one gets the local gauge invariance
anomalies which should cancel to have a consistent theory.

Calculation of the
triangle diagram is the same for all anomalies, the only difference
being described by the group-theoretical factor. The triangle diagram
$\langle j^{a, \mu}_G j^{b, \nu}_G j^{c, \lambda}_{G}\rangle$,
where $j_G^{a, \mu}=\bar{\psi} \gamma^\mu t^a \psi$
with $\psi$ being the fermion component of a chiral
superfield, is proportional to $\mathcal{A}(\mathcal{R})
 = \textup{Tr} [t^a \{t^b,t^c\}]$ (the trace is taken over $G_c$-group
matrices in some representation $\mathcal{R}$).
The total anomaly is proportional to the sum over all fermions
$\sum_{\text{fermions}} \textup{Tr} [t^a \{t^b,t^c\}]$.
In our electric theory the explicit calculation boils to the equality
\beq \label{gauge_an}
(1) N_f + (-1) N_f + 0 = 0,
\eeq
since the triple Casimir invariant of $G_c=SU(N_c)$ for the fundamental
representation is $\mathcal{A}(f)=1$, for the antifundamental one
$\mathcal{A}(\overline{f})=-1$, and for the adjoint representation
$\mathcal{A}(adj)=0$.

In the definition of SCIs \eqref{Ind} it is assumed that all operators
entering it represent exact physical symmetries.
This means that the corresponding currents are not anomalous
$\langle j^{a, \mu}_G j^{b, \nu}_G j^{\lambda}_1\rangle=0$,
where $j_1^\lambda$ is $U(1)_R$ or any flavor symmetry current
(in the infrared fixed point $R$-charge should be
conserved similar to the energy-momentum).
In the $U(1)_R$-case this anomaly coefficient is proportional to
$R \textup{Tr} \{t^a t^b\} = R T(\mathcal{R}) \delta^{ab}$,
where $R$ is the $R$-charge. In the electric theory one has
\beq \label{R_eq}
(R-1) 2 N_f \frac 12 +  N_c = 0,
\eeq
since the Casimir operators of $G_c=SU(N_c)$ for fundamental,
antifundamental, and adjoint representations are
$T(f)=T(\overline{f})=1/2$, and $T(adj)=N_c$, respectively.
Here $R-1$ is the $R$-charge of chiral quarks and the $R$-charge of
gluinos is equal to 1. As a result, one fixes the $R$-charge of chiral superfields,
$R=(N_f-N_c)/N_f$. Similarly, gauge invariance yields
$\langle j^{a, \mu}_G j^{\nu}_1 j^{\lambda}_2\rangle=0$
for any conserved global symmetry current $j^{\nu}_1$ and $j^{\lambda}_2$.

As to the anomalies associated only with global symmetry groups --
they are not obliged to vanish. As argued by 't Hooft \cite{Hooft},
for any electric-magnetic duality (including the Seiberg $\mathcal{N}=1$ duality)
the coefficients of admissible triangle anomalies should
match in dual theories. For example, in the Seiberg case
$SU(N_f)_l^3$-anomaly is described by
$\langle j^{a, \mu}_{SU(N_f)_l} j^{b, \nu}_{SU(N_f)_l} j^{c ,\lambda}_{SU(N_f)_l}\rangle$
with $j^{a, \mu}_{SU(N_f)_l} = \bar{\psi} \gamma^\mu t^a \psi$,
where $t^a$ is the $SU(N_f)_l$ fundamental representation matrix.
For the electric theory the anomaly coefficient comes only from one field and
equals to $(1) N_c$,
while in the magnetic side one has two different contributions
(from dual quarks $q$ and mesons $M$) which yield the coefficient
$(-1) (N_f-N_c) + N_f = N_c$ confirming
one of the 't Hooft anomaly matching conditions.

Being a quantitative check, the 't Hooft anomaly matching conditions
provide an extremely powerful tool for checking $4d$
$\mathcal{N}=1$ dualities. Still, one should be careful with
these conditions, see e.g. \cite{Brodie:1998vv}, where examples
of misleading anomaly matching conditions were found. Namely,
there are $\mathcal{N}=1$ SYM theories with equal anomaly
coefficients, but the deformation by mass parameters argument
shows that these theories are not dual to each other.
From the SCI point of view this fact is reflected in
the difference of analytical structure of SCIs \cite{V}.

We consider explicitly only the original Seiberg duality assuming that
other dualities can be treated in a similar way.
For a further comparison we give a full list of corresponding
nontrivial anomaly coefficients:
 \beqa
SU(N_f)_{l,r}^3 &:& \quad N_c,
\qquad SU(N_f)_{l,r}^2 U(1)_B \ : \qquad \frac{N_c}{2},
\nonumber \\ SU(N_f)_{l,r}^2 U(1)_R &:& \quad (R-1) N_c \frac 12 = -\frac{N_c^2}{2N_f},
\nonumber \\ U(1)_B^2 U(1)_R &:& \quad (R-1) 2 N_f N_c = -2 N_c^2,
\nonumber \\ U(1)_R &:& \quad (R-1) 2 N_f N_c + N_c^2-1 = -N_c^2-1,
\nonumber \\ U(1)_R^3 &:& \quad  (R-1)^3 2 N_f N_c + N_c^2-1 = -2 \frac{N_c^4}{N_f^2}+N_c^2-1,
\eeqa
where $R=(N_f-N_c)/N_f$. Note that
in the case of $U(1)_R$-current anomaly the triangle diagrams involve
two gravitational currents.

{\bf The total ellipticity condition.}
The notion of total ellipticity was introduced first
for elliptic hypergeometric series \cite{S3} which we skip.
An elliptic function is called totally elliptic if it is doubly
periodic in all continuous variables used to parametrize its divisor space
of maximal possible dimension.
A meromorphic function is called the elliptic hypergeometric
term if it satisfies a homogeneous linear difference equation
in one of the variables with the coefficient which is elliptic
in this variable. Elliptic hypergeometric term is called totally elliptic if
it satisfies such equations in each variable with the coefficients
which are totally elliptic functions \cite{spi:cirm}.
It is believed that
one can associate a supersymmetric duality with each nontrivial
totally elliptic hypergeometric term formed as the ratio
of the kernels of two differently looking, but equal integrals \cite{SV2}.

In \cite{spi:cirm} the total ellipticity condition for the equality of
integrals of interest \eqref{IE-seiberg} and \eqref{IM-seiberg}
has been checked. We partially repeat here corresponding calculations.
First, we change variables $\widetilde{\z}$ in
(\ref{IM-seiberg}) to $\widetilde{\z}= S^{-1/\widetilde{N}_c}
\w$. Then the equality of integrals (\ref{IE-seiberg}) and
(\ref{IM-seiberg}) is rewritten in the following way
\begin{eqnarray}\label{SUIntn} &&
\int_{\mathbb{T}^{N_c-1}} \Delta_E(\z,\t,\s)
\prod_{j=1}^{N_c-1} \frac{d z_j}{2 \pi \textup{i}    z_j} =
\int_{(S^{1/\widetilde{N}_c}\mathbb{T})^{\widetilde{N}_c-1}}\Delta_M(\w,\t,\s)
\prod_{j=1}^{\widetilde{N}_c-1} \frac{d w_j}{2 \pi \textup{i} w_j},
\\ \label{kE} && \makebox[-2em]{}
\Delta_E(\z,\s,\t) = \kappa_{N_c} \frac{\prod_{i=1}^{N_f} \prod_{j=1}^{N_c}
\Gamma(s_iz_j,t_i^{-1}z_j^{-1};p,q)}{
\prod_{1 \leq i <   j \leq N_c} \Gamma(z_iz_j^{-1},z_i^{-1}z_j;p,q)},
\\ && \makebox[-2em]{}
\Delta_M(\w,\s,\t) = \kappa_{\widetilde{N}_c}
\prod_{i,j=1}^{N_f} \Gamma(s_it_j^{-1};p,q)
\frac{\prod_{i=1}^{N_f} \prod_{j=1}^{\widetilde{N}_c}
\Gamma(s_i^{-1}w_j,pqt_iw_j^{-1};p,q)}{\prod_{1 \leq i <   j \leq
\widetilde{N}_c} \Gamma(w_iw_j^{-1},w_i^{-1}w_j;p,q)},
\label{kM}\end{eqnarray}
with $\prod_{i=1}^{N_c} z_i=1$ and $\prod_{i=1}^{\widetilde{N}_c}w_i=S.$
Consider the function
\beq \label{ratio}
 \rho(\z,\w,\s,\t;p,q)
\ = \ \frac{\Delta_E(\z,\s,\t)}
{\Delta_M(\w,\s,\t)} \eeq
and the ratios, called $q$-certificates,
\beq\label{hg}
h^{g}(\z,\w,\s,\t,q;p) \ = \ \frac{\rho(\ldots,q g,\ldots;p,q)}
{\rho(\ldots,g,\ldots;p,q)}, \quad g \in \{ \z,\w,\s,\t \}.
\eeq
The total ellipticity condition for $\rho(\z,\w,\s,\t;p,q)$ is then formulated as
the requirement for all $h^{g}$-functions to be $p$-elliptic
in all variables $\{ \z,\w,\s,\t,q \}$, i.e. they should not change
under the $p$-shifts $z_i\to p^{\alpha_i}z_i, w_i\to p^{\beta_i}w_i,
s_k\to p^{\gamma_k}s_k, t_k\to p^{\mu_k}t_k, q\to p^{\nu}q$,
$\alpha_i,\ldots,\nu \in \Z,$ provided all the additional constraints
for fugacities are satisfied.
We remind that according to the original definition given in \cite{S2},
a contour integral with integration variables $z_i$ is called the elliptic
hypergeometric integral if $h^{z_i}$-certificates built from its kernel
are $p$-elliptic in all $\z$, which is a much weaker condition.

Let $m_j^{(a)}\in\Z$, $j=1,\dots,n,\ a=1,\dots, K$, and
$\epsilon(m^{(a)})=\epsilon(m^{(a)}_1,\dots,m^{(a)}_n)$
are arbitrary  $\Z^n\to \Z$ maps with finite support and $r_{-}\in\Z$.
Define a meromorphic function of free variables $x_i\in\mathbb{C}^*,\ i=1,\ldots,n,$
\begin{equation}
\Delta(x_1,\dots,x_n;p,q) = (p;p)_\infty^{r_{-}}(q;q)_\infty^{r_{-}}\prod_{a=1}^K
\Gamma(x_1^{m_1^{(a)}}x_2^{m_2^{(a)}} \dots
x_n^{m_n^{(a)}};p,q)^{\epsilon(m^{(a)})}.
\label{rs-term}\end{equation}
The following theorem was presented in \cite{spi:cirm}.

\begin{theorem} [Rains, Spiridonov, 2004]
Suppose $\Delta(x;p,q)$ is a totally elliptic hypergeometric term,
i.e. all its $q$-certificates
$$
h_i(x,q;p)=\frac{\Delta(\ldots, qx_i, \ldots;p,q)}{\Delta(x_1,\dots,x_n;p,q)}
=\prod_{a=1}^K\prod_{l=0}^{m_i^{(a)}-1}
\theta(q^l\prod_{k=1}^n x_k^{m_k^{(a)}};p)^{\epsilon(m^{(a)})}
$$
are $p$-elliptic functions of $q$ and
$x_1,\ldots,x_n$. Then these $q$-certificates are also modular invariant.
\end{theorem}

The statements of the theorem are guaranteed because of the
following diophantine equations
\begin{eqnarray} \label{con1}
&& \sum_{a=1}^K \epsilon(m^{(a)}) m_i^{(a)} m_j^{(a)} m_k^{(a)} = 0,\\
&& \sum_{a=1}^K \epsilon(m^{(a)}) m_i^{(a)} m_j^{(a)}     = 0,
\label{con2}
\\
&& \sum_{a=1}^K \epsilon(m^{(a)}) m_i^{(a)} =0.
 \label{con3}\end{eqnarray}
The proof is elementary. The demand $h_i(\ldots px_j\ldots,q;p)=h_i(x,q;p)$
leads to equations \eqref{con1}, \eqref{con2}. Equation
\eqref{con3} emerges as a consequence of the restriction $h_i(x,pq;p)=h_i(x,q;p)$.
The theorem statement follows after application to each theta
function in $h_i$ the modular transformation \eqref{mod_theta} and use
of equations \eqref{con1}-\eqref{con3}.

In the context of SCIs variables $x_i$ represent combinations of chemical potentials
of symmetry groups, $\Delta$-function is the ratio of kernels of dual indices,
 and $r_{-}$ is the difference of ranks of the electric
and magnetic gauge groups,
$$
r_{-}=r_e-r_m,\quad r_e=\text{rank } G_c,\quad r_m=\text{rank } \widetilde{G}_c.
$$

During the checks of the total ellipticity condition for known dualities
in \cite{SV2} we have
noticed that some phases of the quasiperiodicity factors emerging from $p$-shifts
for contributions coming from electric (or magnetic) theories coincide with
the anomaly coefficients. This observation allowed us to
conjecture that the total ellipticity condition guarantees 't Hooft anomaly
matchings. As will be shown below this is not the case and
the Rains-Spiridonov equations \eqref{con1}-\eqref{con3} do not
describe a complete set of anomaly matchings of the Seiberg-like dual
theories.

It is necessary to verify that the elliptic hypergeometric term
\eqref{ratio} belongs to the class \eqref{rs-term}, which is not evident
from its explicit form we have given. In order to see this
one should take definitions \eqref{kE} and \eqref{kM},
replace there $t_i\to (pq)^{-1}t_i$ for $i=1,\ldots, N_f-N_c$
(to remove $pq$ from the balancing condition),
and apply the reflection formula $\Gamma(pqz;p,q)=1/\Gamma(z^{-1};p,q)$
to elliptic gamma functions
having the product $pq$ in their arguments.

We stress that the ansatz \eqref{rs-term}
does not describe all possible forms of the elliptic hypergeometric terms.
In general one can have in the arguments of elliptic gamma functions
the non-removable factors $(pq)^{R/2}$ for some fractional numbers $R$
(e.g., this is so for the Kutasov-Schwimmer duality \cite{KS}) in which
case the total ellipticity condition should be modified
appropriately \cite{spi:cirm,SV2}.

{\bf Ellipticity of certificates for $z_i$ and gauge anomalies.}
Take the $q$-certificates for integration variables $\z$
obtained from \eqref{ratio} after rescaling $z_{i} \rightarrow
qz_{i}$ for $i \neq {N_c}$ and $z_{N_c} \rightarrow q^{-1} z_{N_c}$
(i.e., we assume that $z_{N_c}= \prod_{i=1}^{Nc-1} z_i^{-1}$)
and written in terms of the initial variables \eqref{ini_var}:
\begin{eqnarray}\nonumber
&& h^{z}_{i}(\z,v,\x,\y,q) = \frac{\rho(\z,\w,\s,\t;p,q)|_{z_{i}
\rightarrow qz_{i}, z_{N_c} \rightarrow q^{-1} z_{N_c}}}
{\rho(\z,\w,\s,\t;p,q)}
\\ \makebox[-3em]{} && = \frac{ \T(q^{-2}z_{i}^{-1}z_{N_c},q^{-1}z_{i}^{-1}z_{N_c};p)}{
\T(qz_{i}z_{N_c}^{-1},z_{i}z_{N_c}^{-1};p)}
\prod_{j=1,j \neq {i}}^{N_c-1}
 \frac{\T(q^{-1}z_{i}^{-1}z_j,q^{-1}z_j^{-1}z_{N_c};p)}
{\T(z_{i}z_j^{-1},z_jz_{N_c}^{-1};p)} \nonumber
\\ && \makebox[2em]{} \times
\prod_{k=1}^{N_f}
\frac{\T((pq)^{R/2} v x_k z_{i}, (pq)^{R/2} v^{-1} y_k^{-1}
z_{N_c}^{-1};p)}{\T((pq)^{R/2} v^{-1} y_k^{-1} (qz_{i})^{-1},
(pq)^{R/2} v x_kq^{-1}z_{N_c};p)}.
\label{el_ker_SD}\end{eqnarray}
From the physical point of view,
consideration of the $z_i$-variable certificate can be interpreted as the insertion
of one gauge current $j_G^{i,\mu}$ into the triangle anomaly diagram.
In terms of equations \eqref{con1}-\eqref{con3} it means that we deal
with their subpart depending at least linearly on $m_i^{(a)}$ coming from
$G_c$-fugacities.
Since the dependence on $z_i$ in $\rho$ comes only from $\Delta_E$, the same result
is obtained if we replace in \eqref{el_ker_SD} $\rho$ by the kernel of integral
describing electric SCI, i.e. the properties of $h^{z}_{i}$ describe only
the electric theory. Similar situation holds for $w_i$-variables
associated only with the magnetic theory.

It is easy to check that $h^{z}_{i}(\z,v,\x,\y,q)$ is a totally
$p$-elliptic function:
\begin{eqnarray}\nonumber
&& \makebox[-2,5em]{}
\frac{h^{z}_{i}(z_1,\ldots,pz_{i},\ldots,p^{-1}z_{N_c},v,\x,\y,q)}{h^{z}_{i}(\z,v,\x,\y,q)}
= \frac{(pq)^{2(N_f-N_c)}}{(pq)^{2RN_f}\prod_{i=1}^{N_f}x_i^2y_i^{-2}}=1,
\\ && \makebox[-2em]{}
\frac{h^{z}_{i}(z_1,\ldots,pz_{\K},\ldots,p^{-1}z_{N_c},v,\x,\y,q)}{h^{z}_{i}(\z,v,\x,\y,q)}
= \frac{(pq)^{N_f-N_c}}{(pq)^{RN_f}\prod_{i=1}^{N_f}x_iy_i^{-1}}=1,
 \label{listH}\\ && \makebox[-2,5em]{}
\frac{h^{z}_{i}(\z,v,\ldots,px_{\J},\ldots,p^{-1}x_{N_f},\y,q)}
{h^{z}_{i}(\z,v,\x,\y,q)}=
\frac{h^{z}_{i}(\z,v,\x,\ldots,py_{\J},\ldots,p^{-1}y_{N_f},q)}
{h^{z}_{i}(\z,v,\x,\y,q)}=1,
\label{sunf} \\ &&   \makebox[10em]{}
\frac{h^{z}_{i}(\z,pv,\x,\y,q)}
{h^{z}_{i}(\z,v,\x,\y,q)}=1,
\label{u1b} \end{eqnarray}
where $\K \neq {i}$. The most complicate looking identity is
\begin{eqnarray} &&
\frac{h^{z}_{i}(\z,p^{R/2}v,p^{R(N_f-1)}x_1,p^{-R}x_2,\ldots,p^{-R}x_{N_f},
\y,pq)} {h^{z}_{i}(\z,v,\x,\y,q)}=1,
\label{jG}\end{eqnarray}
and its obvious partners obtained by permutation of $x_j$
together with similar equations involving $y_j$-variables.
In terms of the variables $s_j, t_j$ this
symmetry looks more compact: one has the transformations
$s_a\to p^{N_f-N_c} s_a$
(or $t_a\to p^{N_c-N_f} t_a$) for one fixed value of $a$ and $q\to pq$.

If one takes an arbitrary ratio of elliptic
gamma functions whose arguments are given by integer powers of the
fugacities $v,z_i,x_j,y_j$, then the $q$-certificates will be again
given by ratios of theta-functions. However, $p$-shifts of
the fugacities in these certificates would produce in general
arbitrary quasiperiodicity factors described by some
powers of all fugacities (which are all equal to 1 in our case).

Equations (\ref{listH}) fix the second current to be again
the gauge current since we are taking $p$-shifts for the $z_j$-variable
and the resulting quasiperiodic factor phases will necessarily contain
$m_j^{(a)}$-power. The third current in the triangle anomaly is
fixed by considering in the resulting phase the powers of
fugacities $v$ (for the $U(1)_B$-current), $x_k$ (for the $SU(N_f)_l$-current),
$y_k$ (for the $SU(N_f)_l$-current) and for obtaining insertion of
the $U(1)_R$-current one should trace the powers $(pq)^{R/2}$.

Let us pick up cubic products
of $m_i^{a}$ \eqref{con1} corresponding to the gauge group fugacities
and sum over $a$ -- this sum corresponds to the
anomaly coefficient for $\langle j_G^{i,\mu}j_G^{j,\nu}j_G^{k,\lambda}\rangle$
with color indices $i,j,k$.
It is easy to see that it vanishes, moreover, its pieces coming from
the gluinos (i.e., from the terms $\propto\Gamma(z_iz_j^{-1},z_i^{-1}z_j$)
and the chiral fields vanish independently.
Cancellation of the powers of the $v$-variables
in (\ref{listH}) tells that the gauge anomaly $SU(N_c)^2U(1)_B$ is absent,
and similar situation holds for $SU(N_c)^2 SU(N_f)_{l,r}$-anomalies.

If the $R$-charge is not fixed in advance, then there emerge
quasiperiodicity multipliers given by some powers of $pq$, as indicated
in \eqref{listH}. The demand of the absence of these multipliers
fixes the $R$-charge in the
same way as the vanishing of gauge anomaly
$\langle j^{a \mu}_G j^{b \nu}_G j^{\rho}_{U(1)_R}\rangle=0$
does, $N_f-N_c-RN_f=0$. Absence of the asymmetry in $p$ and $q$ in these multipliers,
despite of such asymmetry present in \eqref{el_ker_SD}, corresponds to the energy-momentum
conservation.

Equations \eqref{sunf} correspond to the choice of the second
current in the anomaly triangle diagram as $SU(N_f)_{l,r}$-currents
since we scale respective fugacities. Then the third current
is determined from the quasiperiodicity factors. Absence of
such factors in our case shows that all corresponding anomalies vanish.
Thus, separate vanishing of polynomials \eqref{con1}-\eqref{con3}
for electric and magnetic theories,
when at least one of $m_i^{(a)}$-variables comes from
gauge group fugacities, describes cancellation of gauge anomalies
and various conservation laws.

One can consider in a similar way other certificates and interpret
corresponding ellipticity constraints as anomaly matching conditions,
but this construction is not that lucid and evident as one would want to.
Moreover, since there is no separate
fugacity for $U(1)_R$-group, there is no $q$-certificate
associated with this group which would correspond to the insertion of
$U(1)_R$-current alone. Therefore, it is not possible to describe $U(1)_R$
and $U(1)_R^3$ anomalies in this way. Similar conclusion has been
reached recently by Sudano \cite{sudano} following our considerations
in \cite{spi:cirm,SV2}. Let us show that the $SL(3,\Z)$-modular properties
of elliptic hypergeometric terms produce {\em all} anomaly matching
conditions at once in a very simple way.

\section{$SL(3,\mathbb{Z})$-Modularity and anomalies}

In \cite{unit} the modified versions of elliptic hypergeometric
integrals have been introduced. They satisfy the general definition
of elliptic hypergeometric integrals of \cite{S2}
mentioned above, but they are built from the modified
elliptic gamma functions. Consider modifications
of integrals \eqref{IE-seiberg}  and \eqref{IM-seiberg}.
For this introduce new parametrization of fugacities
\begin{eqnarray} \nonumber&&
z_j = e^{2 \pi \textup{i} u_j/\omega_2},\  j =1,\ldots,N_c,
\quad \widetilde z_j = e^{2 \pi \textup{i} v_j/\omega_2}, \  j =1,\ldots,\widetilde{N}_c,
\\ &&
s_i = e^{2 \pi \textup{i} \alpha_i/\omega_2},
\ \ \ t_i = e^{2 \pi \textup{i} \beta_i/\omega_2},\  i =1,\ldots,N_f.
\label{newpar}\end{eqnarray}
Define now the following integrals
\begin{eqnarray}\label{el1}
&& I_E^{mod} =  \kappa^{mod}_{N_c} \int_{-\omega_3/2}^{\omega_3/2}
\frac{\prod_{i=1}^{N_f} \prod_{j=1}^{N_c}
\mathcal{G}(\alpha_i+u_j, -\beta_i-u_j;\mathbb{\omega})}
{\prod_{1 \leq i <   j \leq N_c} \mathcal{G}(u_i-u_j,-u_i+u_j;\mathbb{\omega})}
 \prod_{j=1}^{N_c-1} \frac{du_j}{\omega_3},
\end{eqnarray}
where $\sum_{j=1}^{N_c} u_j =0$,
$\mathcal{G}(a,b;\mathbb{\omega}):=\mathcal{G}(a;\mathbb{\omega})
\mathcal{G}(b;\mathbb{\omega})$, and
\begin{eqnarray}\label{mag1}
&& I_M^{mod} =  \kappa^{mod}_{\widetilde{N}_c} \prod_{1 \leq i,j \leq N_f}
\mathcal{G}(\alpha_i-\beta_j;\mathbb{\omega})
\\ \nonumber && \makebox[3em]{} \times \int_{-\omega_3/2}^{\omega_3/2}
\frac{\prod_{i=1}^{N_f} \prod_{j=1}^{\widetilde{N}_c}
\mathcal{G}(\alpha/\widetilde{N}_c -\alpha_i+v_j,
-\beta/\widetilde{N}_c+\beta_i-v_j;\mathbb{\omega})}
{\prod_{1 \leq i <   j \leq \widetilde{N}_c}
\mathcal{G}(v_i-v_j, -v_i+v_j;\mathbb{\omega})} \prod_{j=1}^{\widetilde{N}_c-1}
\frac{dv_j}{\omega_3},
\end{eqnarray}
where $\widetilde{N}_c=N_f-N_c$ and $\sum_{j=1}^{\widetilde{N}_c}v_j=0$.
The integration in both cases goes along the straight line segment
connecting points $-\omega_3/2$ and $\omega_3/2$.
The balancing condition reads
$$
\alpha-\beta = (N_f-N_c) \sum_{k=1}^3\omega_k,
\qquad \alpha =\sum_{i=1}^{N_f} \alpha_i, \quad
\beta =\sum_{i=1}^{N_f} \beta_i.
$$
Finally,
$$
\kappa_{N_c}^{mod} = \frac{\kappa(\mathbf{\omega})^{N_c-1}}{N_c!}, \qquad
\kappa(\mathbf{\omega}) =-\frac{\omega_3}{\omega_2}
\frac{(p;p)_\infty(q;q)_\infty(r;r)_\infty}
{(\tilde{q};\tilde{q})_\infty}.
$$

These integrals are obtained from \eqref{IE-seiberg}  and \eqref{IM-seiberg}
after replacement of $\Gamma(z;p,q)$ with $z=e^{2\pi\textup{i}u/\omega_2}$
by the function $\mathcal{G}(u;\mathbb{\omega})$ and some modification
of the integration measure. Since both elliptic gamma functions
solve the key equation \eqref{eq1}, the modified elliptic hypergeometric integrals
satisfy the same finite difference equations in the shifts $u\to u+\omega_1$
as the standard integrals do (and therefore modified identities can be
proved similarly to the standard ones). However, they remain well defined for $|q|=1$ in
difference from integrals \eqref{IE-seiberg}  and \eqref{IM-seiberg}.

\begin{theorem} Suppose that
$$
\Im(\alpha_i/\omega_3),\Im((\alpha/\widetilde N_c-\alpha_i)/\omega_3)  <0, \quad
\Im(\beta_i/\omega_3), \Im((\beta/\widetilde N_c-\beta_i)/\omega_3)>0.
$$
Then $I_E^{mod}=I_M^{mod}.$
\end{theorem}

The simplest proof follows the same lines as in \cite{unit}, where a
similar identity has been established for modified elliptic
hypergeometric integrals of type II on the $BC_n$-root system.
Namely, one should substitute to \eqref{el1}, \eqref{mag1}
the modular transformed form of the modified elliptic gamma
function \eqref{MEGF2} and simplify the combination of
$B_{3,3}$-Bernoulli polynomials in the exponential factors.
After application of the modular transformation law
for the Dedekind eta-function \eqref{Ded} to infinite
products $(p;p)_\infty, (q;q)_\infty, (r;r)_\infty$
these multipliers cancel out completely. As a result the
equality $I_E^{mod}=I_M^{mod}$ reduces to the equality $I_E=I_M$
with the variables $s_j, t_j, p$ and $q$ replaced by
$e^{-2 \pi \textup{i} \alpha_j/\omega_3},
e^{-2 \pi \textup{i} \beta_j/\omega_2}, \tilde p$ and $\tilde r$,
respectively. The constraints imposed on the variables $\alpha_j$
and $\beta_j$ convert to the restrictions needed for guaranteeing
the equality of integrals.

Denote as $I_{E,M}(\underline{\alpha},\underline{\beta};
\omega_1,\omega_2,\omega_3)$
the SCIs $I_{E,M}(s,t;p,q)$ with the change of parameters \eqref{newpar}.
Then Theorem 2 states that integrals \eqref{el1} and \eqref{mag1}
are proportional to $SL(3,\Z)$-modular transformations
$(\omega_1,\omega_2,\omega_3) \to (\omega_1,-\omega_3,\omega_2)$
of the original integrals
$$
I_E^{mod}=e^{\varphi_e}I_E(\underline{\alpha},\underline{\beta};
\omega_1,-\omega_3,\omega_2), \quad
I_M^{mod}=e^{\varphi_m}I_M(\underline{\alpha},\underline{\beta};
\omega_1,-\omega_3,\omega_2),
$$
and $\varphi_e=\varphi_m$. The latter equality appears to be nothing
else than the 't Hooft anomaly matching conditions!
Let us prove this statement in the general setting.

Given arbitrary  $\Z^n\to \Z$ maps with finite support $m_j^{(a)}\in\Z$,
$j=1,\dots,n$, $\epsilon(m^{(a)})=\epsilon(m^{(a)}_1,
\dots,m^{(a)}_n), \ a=1,\dots, K$, some $r_{-}\in\Z$ and
real numbers $R^{(a)}\in\mathbb{R}$, we define a meromorphic function of
$x_i\in\mathbb{C}^*,\ i=1,\ldots,n,$
\begin{equation}
\Delta(x_1,\dots,x_n;p,q) = (p;p)_\infty^{r_{-}}(q;q)_\infty^{r_{-}}\prod_{a=1}^K
\Gamma\Bigl((pq)^{\frac{R^{(a)}}{2}}x_1^{m_1^{(a)}}x_2^{m_2^{(a)}} \dots
x_n^{m_n^{(a)}};p,q\Bigr)^{\epsilon(m^{(a)})}.
\label{gen-term}\end{equation}
One can demand that the powers of $pq$ are not removable
by the transformations $x_j\to (pq)^{\gamma_j}x_j$,
i.e. that there do not exist real numbers $\gamma_j$ such that
$R^{(a)}/2+\sum_{j=1}^n\gamma_jm_j^{(a)}=0$. However, we shall
not require this for simplicity.

Denote now $x_j=e^{2\pi\textup{i}u_j/\omega_2}$ and introduce
the following meromorphic function of $u_j\in\mathbb{C}$:
\begin{equation}
\Delta^{mod}(u_1,\dots,u_n;\mathbf{\omega}) =  \kappa(\mathbf{\omega})^{r_{-}}
\prod_{a=1}^K \mathcal{G}\Bigl(R^{(a)}\sum_{k=1}^3\frac{\omega_k}{2}
+\sum_{j=1}^nu_jm_j^{(a)};\mathbf{\omega}\Bigr)^{\epsilon(m^{(a)})}.
\label{gen-modterm}\end{equation}
The modular transformation properties of the totally elliptic
hypergeometric terms were investigated in \cite{spi:cirm}. In the
present context we have the following theorem.

\begin{theorem} The $SL(3,\Z)$-modular transformation invariance relation
\begin{equation}
\Delta^{mod}(u_1,\dots,u_n;\mathbf{\omega}) = \Delta(e^{-2\pi\textup{i}u_1/\omega_3},
\ldots,e^{-2\pi\textup{i}u_n/\omega_3};\tilde p,\tilde r)
\label{modularity}\end{equation}
leads to the following set of equations
\begin{eqnarray}\label{mcon1}
&& \sum_{a=1}^K \epsilon(m^{(a)}) m_i^{(a)} m_j^{(a)} m_k^{(a)} = 0,\\
&& \sum_{a=1}^K \epsilon(m^{(a)}) m_i^{(a)} m_j^{(a)}(R^{(a)}-1)= 0,
\label{mcon2} \\
&& \sum_{a=1}^K \epsilon(m^{(a)}) m_i^{(a)}(R^{(a)}-1)^2 =0,
 \label{mcon3}
\\
&& \sum_{a=1}^K \epsilon(m^{(a)}) m_i^{(a)} =0,
 \label{mcon4}
\\
&& \sum_{a=1}^K \epsilon(m^{(a)}) (R^{(a)}-1)^3+r_{-}=0,
 \label{mcon5}
\\
&& \sum_{a=1}^K \epsilon(m^{(a)}) (R^{(a)}-1)+r_{-} =0.
 \label{mcon6}
\end{eqnarray}
\end{theorem}

The proof is simple enough. From representation \eqref{MEGF2}
is it easy to see that
\begin{eqnarray*} &&
\frac{\Delta^{mod}}{\Delta}=\frac{\kappa(\mathbf{\omega})^{r_{-}}}
{(p;p)_\infty^{r_{-}}(q;q)_\infty^{r_{-}} }
\prod_{a=1}^K\exp\Bigg[-\frac{\pi\textup{i}\epsilon(m^{(a)})}
{3\omega_1\omega_2\omega_3}\Bigl(
\frac{R^{(a)}-1}{2}\sum_{k=1}^3\omega_k+\sum_{i=1}^n u_im_i^{(a)}\Bigr)
\\ && \makebox[4em]{} \times
\Bigl( \Big(\frac{R^{(a)}-1}{2}\sum_{k=1}^3\omega_k+\sum_{i=1}^nu_im_i^{(a)}\Big)^2
-\frac{1}{4}\sum_{k=1}^3\omega_k^2\Bigr) \Bigg]=1.
\end{eqnarray*}
Since chemical potentials $u_i$ are continuous independent variables,
the polynomial in the exponential depending on them should vanish.
Picking up the cubic terms $u_iu_ju_k$ one obtains equation \eqref{mcon1},
the quadratic terms yield \eqref{mcon2}, there are two terms linear in $u_i$
with the coefficients depending on continuous modular parameters $\omega_k$
in different way. Vanishing of these
terms yields two equations \eqref{mcon3} and \eqref{mcon4}.
Finally, we are left with the equation
\begin{eqnarray*} &&
\Biggl[-\frac{\omega_3(p;p)_\infty(q;q)_\infty(r;r)_\infty}
{\omega_2(\tilde p;\tilde p)_\infty(\tilde q;\tilde q)_\infty
(\tilde r;\tilde r)_\infty}\Biggr]^{r_{-}}
\prod_{a=1}^K\exp\Biggl[-\frac{\pi\textup{i}\epsilon(m^{(a)})}
{24\omega_1\omega_2\omega_3}(\sum_{k=1}^3\omega_k)
\\  && \makebox[2em]{} \times
(R^{(a)}-1)\Bigl((R^{(a)}-1)^2(\sum_{k=1}^3\omega_k)^2
-\sum_{k=1}^3\omega_k^2\Bigr)\Biggr]=1.
\end{eqnarray*}
Applying the modular transformation formula \eqref{Ded} to infinite products
and using arbitrariness of variables $\omega_k$ we come
to the last two equations \eqref{mcon5} and \eqref{mcon6}.

Suppose now that the powers $(pq)^{R^{(a)}/2}$ can be removed from
\eqref{gen-term} by redefinition of variables $x_i\to (pq)^{\gamma_i}x_i$,
i.e. that there exist some numbers $\gamma_i$ such that
$R^{(a)}=-2\sum_{i=1}^n\gamma_im_i^{(a)}$.
Substituting these conditions to \eqref{mcon2}, \eqref{mcon3}, we immediately see
that they reduce to equations \eqref{con2}, \eqref{con3}, i.e. the situation
becomes similar to the original Seiberg duality case.
Interestingly, equations \eqref{mcon5} and \eqref{mcon6} are reduced
in this case to one constraint
\begin{equation}
\sum_{a=1}^K \epsilon(m^{(a)}) =r_{-}.
\label{mcon6'}\end{equation}
If the ranks of dual gauge groups are equal (e.g., for self-dual theories),
one has $\sum_{a=1}^K \epsilon(m^{(a)}) =0.$ Equation \eqref{mcon6'}
thus completes equations \eqref{con1}-\eqref{con3} to guarantee $SL(3,\Z)$-modular
invariance of such elliptic hypergeometric terms \cite{spi:cirm}.

It is evident that equations \eqref{mcon1}-\eqref{mcon6}
coincide with the 't Hooft anomaly matching conditions
for dual theories with the $\Delta$-function being built as the ratio of kernels of
elliptic hypergeometric integrals describing electric and magnetic SCIs.
We have checked this statement explicitly for the original
Seiberg duality using the kernels of modified elliptic hypergeometric
integrals \eqref{el1} and \eqref{mag1} with the substitutions
\begin{eqnarray*} &&
\alpha_i = R(\omega_1+\omega_2+\omega_3)/2 + \eta + \delta_i,
\\ &&
\beta_i = -R(\omega_1+\omega_2+\omega_3)/2 + \eta + \xi_i, \quad i=1,\ldots,N_f,
\end{eqnarray*}
where $\eta$ is the chemical potential for $U(1)_B$-group, $\delta_i$ and
$\xi_i$ are chemical potentials for $SU(N_f)_l$ and $SU(N_f)_r$ groups,
$\sum_{i=1}^{N_f}\delta_i=\sum_{i=1}^{N_f}\xi_i=0$.
The general rule of getting the anomaly coefficients is very simple:
expand $SL(3,\mathbb{Z})$-phase factor and associate the gauge and flavor group
currents with the corresponding chemical potentials and the $U(1)_R$-current
with the term $R^{(a)}-1$, describing the $R$-charge of the chiral
fermions and for $R^{(a)}=0$ modelling the contribution of gluinos.
Since we have a cubic polynomial in these variables we model the
triangle anomaly diagram. For instance, the plain chiral superfield
contributes to the modular phase the term
$\propto B_{3,3}(R(\omega_1+\omega_2+\omega_3)/2;\mathbf{\omega})$
which is easily seen to contain only two pieces $\propto (R-1)^3$ and
$\propto (R-1)$, as needed for $U(1)^3_R$ and $U(1)_R$-anomalies.

Computing the modular transformation exponential factors
for the electric theory alone we explicitly see
emergence of {\em all} anomaly coefficients (coinciding with the
magnetic theory coefficients):

\begin{itemize}
 \item
Cubic polynomials depending on the integration variables $u_i$ or $v_i$
lead to equations \eqref{mcon1}-\eqref{mcon4} with at least one index $i$
coming from the gauge groups. They vanish separately in electric and
magnetic theories (this is true for any duality, not just the Seiberg case)
leading to $\langle j^{a, \mu}_G j^{\nu}_1 j^{\lambda}_2\rangle$=0
for any conserved current $j^{\lambda}_{1,2}$ including the energy momentum tensor.
E.g., from equation \eqref{mcon2} one finds $R=(N_f-N_c)/N_f$.

\item The terms $\propto \delta_i\delta_j\delta_k$ corresponding
to \eqref{mcon1} yield the $SU(N_f)_l^3$-anomaly
coefficient $\propto N_c$ (with a similar result for $SU(N_f)_r^3$).

\item The terms $\propto \delta_i\delta_j\eta$ corresponding to
\eqref{mcon1} give the $SU(N_f)_l^2 U(1)_B$-anomaly
coefficient $\propto N_c$.

\item The terms $\propto \delta_i\delta_j (R-1)$ corresponding to
\eqref{mcon2} give the $SU(N_f)_l^2 U(1)_R$-anomaly
coefficient $\propto N_c^2/N_f$.

\item The terms $\propto \eta^2(R-1)$ corresponding to
\eqref{mcon2} give the $U(1)_B^2 U(1)_R$-anomaly coefficient $\propto N_c^2$.

\item The terms $\propto (R-1)^2$ corresponding to \eqref{mcon3}
are absent leading to vanishing $U(1)_BU(1)_R^2$
and $SU(N_f)_{l,r}U(1)_R^2$-anomalies.

\item Linear terms in flavor chemical potentials are absent
(i.e., equations \eqref{mcon4} are satisfied separately for electric and magnetic
theories), which means that there are no $U(1)_B$ and $SU(N_f)_{l,r}$-anomalies.

\item The electric part of equation \eqref{mcon5} yields precisely the $U(1)_R^3$-anomaly
coefficient $2N_fN_c(R-1)^3+N_c(N_c-1)+\text{rank }G_c=-2N_c^4/N_f+N_c^2-1$.

\item The electric part of equation \eqref{mcon6} yields precisely the $U(1)_R$-anomaly
coefficient $2N_fN_c(R-1)+N_c(N_c-1)+\text{rank }G_c=-N_c^2-1$.

\end{itemize}

If $\Im (\omega_1/\omega_2)>0$ then one can take the limit
$\omega_3 \rightarrow \infty$ and obtain
\beq
\lim_{\omega_3 \rightarrow \infty} \mathcal{G}(u;\mathbb{\omega})
=\frac{(e^{2 \pi \textup{i} u/\omega_1}\widetilde{q};\widetilde{q})_\infty}
{(e^{2 \pi \textup{i} u/\omega_2};q)_\infty}.
\label{4d3d}\eeq
Taking this limit in the relation $I_E^{mod}=I_M^{mod}$ one gets the equality
of partition functions of some $3d$ $\mathcal{N}=2$ theories which
is similar to the reduction of standard $4d$ SCIs to $3d$ partition
functions \cite{DSV}. The main difficulty in finding $3d$ Seiberg dualities
consists in the absence of the anomaly matching conditions.
Starting from known $4d$ dualities and using the limit \eqref{4d3d}
one automatically and easily derives $3d$ dual theories which comprise
(in a hidden way) $4d$ anomaly matching conditions.

\section{Total ellipticity and modularity. The general case.}

Let us consider the total ellipticity condition
for general elliptic hypergeometric term \eqref{gen-term}.
Corresponding $q$-certificates have the form
\begin{equation}
h_i(x,q;p)=\prod_{a=1}^K\prod_{l=0}^{m_i^{(a)}-1}
\theta\Big(q^l(pq)^{R^{(a)}/2}\prod_{k=1}^n x_k^{m_k^{(a)}};p\Big)^{\epsilon(m^{(a)})}.
\label{gen-cert}\end{equation}
Recursively using relation \eqref{quasi} one can verify that the
condition $h_i(\ldots px_j\ldots,q;p)=h_i(x,q;p)$ yields
equations \eqref{mcon1} and \eqref{mcon2} together with the
constraint
\begin{equation}
\sum_{a=1}^K \epsilon(m^{(a)}) m_i^{(a)} m_j^{(a)}\in 2\Z
 \label{mcon2'}\end{equation}
coming from the positive sign prescription.

In order to investigate $p$-periodicity properties of $h_i(x,q;p)$
\eqref{gen-cert} it is
necessary to introduce a new parameter $L$, a minimal positive
integer such that all $LR^{(a)}\in 2\Z$. Note that this
requires an advance knowledge of some properties of $R$-charges,
which are presumed to be fixed from the anomaly cancellation/matching
conditions. Therefore the constraint $h_i(x,p^Lq;p)=h_i(x,q;p)$ looks
a little bit unnatural from the physical point of view.
Nevertheless, it yields equations \eqref{mcon3} and \eqref{mcon4}
together with the constraint
\begin{equation}
L\sum_{a=1}^K \sum_{a=1}^K \epsilon(m^{(a)}) m_i^{(a)}(m_i^{(a)}+R^{(a)})
\in 4\Z.
\label{mcon4'}\end{equation}

Equations \eqref{mcon1}-\eqref{mcon4} and the prescriptions \eqref{mcon2'},
\eqref{mcon4'} were derived from the total ellipticity condition
by the first author (unpublished) in a slightly
different form during the work on \cite{spi:cirm} and more recently by Sudano
in \cite{sudano} (where one can find the details of computations).

On the one hand, both equations \eqref{mcon2'} and \eqref{mcon4'}
do not emerge from the $SL(3,\Z)$-modular invariance condition \eqref{modularity}.
On the other hand, checks of the total ellipticity condition
for all known dualities performed in \cite{SV2} show that they
are satisfied in physical theories. In some cases it can be
shown that they follow from equations \eqref{mcon1}-\eqref{mcon6}
(e.g., for $R^{(a)}\propto \sum_{i=1}^n\gamma_im_i^{(a)}$).
Therefore we conjecture that  equations
\eqref{mcon2'} and \eqref{mcon4'} are automatically
satisfied for elliptic hypergeometric integrals associated with
nontrivial electric-magnetic dualities. If it were true, one
could state that the condition of total ellipticity of elliptic
hypergeometric terms is necessary, but not sufficient for
guaranteeing the 't~Hooft anomaly matching conditions.

Finally, we have a generalization of Theorem 1.

\noindent
{\bf Corollary.}
{\em Suppose  \eqref{gen-term} is a totally elliptic hypergeometric term.
Then all its $q$-certificates $h_i(x,q;p)$ \eqref{gen-cert}
are modular invariant.}
\medskip

For proving this statement consider the ratio of modular transformed
certificates
\begin{eqnarray*} && \makebox[-2em]{}
\frac{\tilde h_i}{h_i}=\prod_{a=1}^K\prod_{\ell=0}^{m_i^{(a)}-1}
\frac{\theta(e^{-2\pi\textup{i}\gamma_\ell^a/\omega_3};\tilde p)^{\epsilon(m^{(a)})} }
{\theta(e^{2\pi\textup{i}\gamma_\ell^a/\omega_2};p)^{\epsilon(m^{(a)})}},
\quad
\gamma_\ell^a=R^{(a)}\sum_{k=1}^3\frac{\omega_k}{2} +\sum_{j=1}^nu_jm_j^{(a)}
+\omega_1\ell.
\end{eqnarray*}
Using the modular transformation law for theta functions \eqref{mod_theta}
one easily checks that $\tilde h_i/h_i=1$ as a consequence
of equations \eqref{mcon1}-\eqref{mcon4}.

\section{Conclusion}

In  \cite{SV2} we formulated the conjecture that all 't Hooft anomaly matching
conditions follow from the total ellipticity condition \cite{spi:cirm}.
It was based on the observation that some of the anomaly coefficients
emerge from the nontrivial quasiperiodicity factors
appearing during the checks of ellipticity of the certificates \eqref{gen-cert}
(in particular, triviality of some factors meant the absence of gauge
anomalies). However, we did not perform a systematic study of this relation
at that time. Later in \cite{DSV} we noticed importance of the $SL(3,\Z)$-modularity
properties for this problem.

In this work we presented a systematic
derivation of the triangle anomaly coefficients for general theories
out of the $SL(3,\Z)$-group modular transformation properties of the
kernels of dual indices. The generalized Rains-Spiridonov equations
\eqref{mcon1}-\eqref{mcon6} are interpreted as the universal 't Hooft
anomaly matching conditions for $4d$ supersymmetric field theories.
In particular, we explicitly checked emergence of all anomaly coefficients
for the original Seiberg duality.

Still, the general physical meaning of the modular transformation properties
of SCIs remains unknown. It is necessary to find physical derivation of
the modified elliptic hypergeometric integrals as
some kind of modified SCIs. Perhaps they
are related to computing indices in $\mathcal{N}=1$ theories
quantized on  $\mathbb{T}^3 \times \mathbb{R}$.
In \cite{HW1}  $4d$ $\mathcal{N}=4$ SYM theories with
simply laced gauge groups were considered on such a space-time.
One can rewrite all SCIs, in particular, SCIs of $4d$ $\mathcal{N}=4$
SYM theories listed in \cite{SV4}, as some modified elliptic
hypergeometric integrals and try to associate our
$SL(3,\mathbb{Z})$-modular transformations with the natural
$SL(3,\mathbb{Z})$-group action in the taken space-time.

Actually, we have {\it demonstrated} coincidence of anomaly matching
conditions with some mathematical properties of SCIs,
but we did not {\it derive} these conditions from first principles.
A proper mathematical consideration of the problem should use
the formal mathematical definition of anomalies as cocycles
of gauge groups (see, e.g., \cite{RSF,Harvey}) yielding
anomaly matching condition as an equality of Chern classes
of dual theories. This should yield also the proper cohomological
meaning of the modular invariance condition for elliptic hypergeometric
terms. From the physical side, it is necessary to compute the effect of
$SL(3,\Z)$-modular transformation on SCIs and demonstrate explicitly
how the anomaly diagrams emerge in the corresponding changes of SCIs.

From the group-theoretical point of view the anomaly coefficients
are described by certain combination of Casimir invariants. It seems
possible to trace how these invariants emerge in the modular
transformation phase using the group-theoretical information
hidden in the definition of SCIs having the elliptic
hypergeometric terms of a specific form (e.g., $r_{-}$
is fixed from a piece of the characters of adjoint representations
of gauge groups). This should also yield anomaly matching conditions.

We would like to finish by posing an interesting mathematical problem
of describing universal restrictions on powers $m_i^{(a)}$ and $\epsilon(m^{(a)})$ in
the general elliptic hypergeometric term \eqref{gen-term}
which would force this term to become a ratio of two kernels
of SCIs \eqref{Ind_fin}. Equations \eqref{mcon1}-\eqref{mcon6}
are necessary for this, but not sufficient.
Such constraints would provide a powerful mathematical tool for
building physical dualities for supersymmetric field theories.

\medskip

{\bf\em Dedication.} {\em This paper is dedicated to the memory of our
friend and collaborator Francis Dolan.
We got acquainted with him because of his beautiful work with Hugh Osborn
on the connection of superconformal indices with the elliptic hypergeometric
integrals \cite{Dolan:2008qi}. From November 2008 we were exchanging
with him by many e-mails, discussed various
aspects of this interrelation and had vast plans for joint work.
Unfortunately, we were able to write only one joint paper \cite{DSV}.
We shall remember Francis as a good friend and a brilliant scientist
with original ideas, and we miss him much.}

\
{\bf Acknowledgments.}
This work is supported in part by RFBR grant no.  12-01-00242 and
the Heisenberg-Landau program.
The authors are indebted to H. Osborn and V. A. Rubakov for
valuable discussions and helpful remarks.
GV would like to thank BLTP, JINR in
Dubna for hospitality in January 2012 during the workshop ``Classical and Quantum
Integrable Systems" where the results of this paper were presented.


\begin{thebibliography}{99}

\bibitem{Dolan:2008qi}
F.~A.~Dolan and H.~Osborn, \textit{Applications of the
Superconformal Index for Protected Operators and $q$-Hypergeometric
Identities to $\mathcal{N}=1$ Dual Theories}, Nucl. Phys.  {\bf
B818} (2009), 137--178.

\bibitem{Kinney} J. Kinney, J. M. Maldacena, S. Minwalla and S. Raju,
\textit{An index for 4 dimensional super conformal theories},
Commun. Math. Phys. {\bf 275} (2007), 209--254.

\bibitem{Romelsberger1} C. R\"omelsberger, \textit{Counting chiral
primaries in ${\mathcal N}=1$, $d=4$ superconformal field theories},
Nucl. Phys. {\bf  B747} (2006), 329--353.

\bibitem{Romelsberger2} C. R\"omelsberger,
\textit{Calculating the superconformal index and Seiberg duality},
{\tt  arXiv:0707.3702 [hep-th]}.

\bibitem{S1}  V. P. Spiridonov,
\textit{On the elliptic beta function}, Uspekhi Mat. Nauk {\bf 56}
(1) (2001), 181--182 (Russian Math. Surveys {\bf 56} (1) (2001),
185--186).

\bibitem{S2} V. P. Spiridonov, \textit{Theta hypergeometric
integrals}, Algebra i Analiz {\bf 15} (6) (2003), 161--215 (St.
Petersburg Math. J. {\bf 15} (6) (2004), 929--967); {\tt math.CA/0303205}.

\bibitem{S3} V. P. Spiridonov, \textit{Essays on the theory of
elliptic hypergeometric functions}, Uspekhi Mat. Nauk {\bf 63} (3)
(2008), 3--72 (Russian Math. Surveys {\bf 63} (3) (2008), 405--472);
{\tt arXiv:0805.3135 [math.CA]}.

\bibitem{Seiberg} N. Seiberg,  \textit{Electric--magnetic duality in
supersymmetric non-Abelian gauge theories}, Nucl. Phys. {\bf B435}
(1995), 129--146.

\bibitem{SV1} V. P. Spiridonov and G. S. Vartanov, \textit{Superconformal
indices for ${\mathcal N}=1$ theories with multiple duals}, Nucl.
Phys. {\bf B824} (2010), 192--216.

\bibitem{SV2}
V.~P.~Spiridonov and G.~S.~Vartanov, \textit{Elliptic hypergeometry
of supersymmetric dualities}, Commun. Math. Phys. {\bf 304} (2011), 797--874.

\bibitem{SV3} V. P. Spiridonov and G. S. Vartanov, \textit{Supersymmetric
dualities beyond the conformal window}, Phys. Rev. Lett. {\bf 105}
(2010) 061603.

\bibitem{SV4}
V.~P.~Spiridonov and G.~S.~Vartanov, \textit{Superconformal indices
of ${\mathcal N}=4$ SYM field theories}, Lett. Math. Phys. {\bf 100}
(2012), 97--118; {\tt  arXiv:1005.4196 [hep-th]}.

\bibitem{V}
G.~S.~Vartanov, \textit{On the ISS model of dynamical SUSY
breaking}, Phys. Lett.  {\bf B696} (2011), 288--290.

\bibitem{S5}
V. P. Spiridonov, \textit{Elliptic beta integrals
and solvable models of statistical mechanics},
Contemp. Math. {\bf 563} (2012), 181--211; {\tt   arXiv:1011.3798 [hep-th]}.

\bibitem{SV5}
V.~P.~Spiridonov and G.~S.~Vartanov,
\textit{Elliptic hypergeometry of supersymmetric dualities II. Orthogonal groups,
knots, and vortices}, {\tt arXiv:1107.5788 [hep-th]}.

\bibitem{Sen} D. Sen, Supersymmetry in the space-time $\mathbb{R}\times S^3$,
Nucl. Phys. {\bf B284} (1987), 201--233.

\bibitem{FS}G.~Festuccia and N.~Seiberg,
\textit{Rigid Supersymmetric Theories in Curved Superspace},
J. High Energy Phys. {\bf 1106 } (2011) 114.

\bibitem{Demail}  F.~A.~Dolan, an e-mail message to the
authors on 10 April 2011 with unpublished notes on localization
in $\mathcal{N}=1$ theories on $S^3\times\mathbb{R}$.

\bibitem{GPRR1} A. Gadde, E. Pomoni, L. Rastelli and S. S. Razamat,
\textit{ $S$-duality and $2d$ Topological QFT}, JHEP {\bf 03} (2010)
032.

\bibitem{GPRR2} A.~Gadde, L.~Rastelli, S.~S.~Razamat and W.~Yan,
\textit{The Superconformal Index of the $E_6$ SCFT}, JHEP {\bf 08}
(2010) 107.

\bibitem{Gadde:2011ik}
A.~Gadde, L.~Rastelli, S.~S.~Razamat and W.~Yan,
\textit{The $4d$ Superconformal Index from $q$-deformed $2d$ Yang-Mills},
Phys. Rev. Lett. {\bf 106} (2011) 241602.

\bibitem{Gadde:2011uv}
A.~Gadde, L.~Rastelli, S.~S.~Razamat and W.~Yan,
\textit{Gauge Theories and Macdonald Polynomials}, {\tt arXiv:1110.3740 [hep-th]}.

\bibitem{Zwiebel:2011wa}
B.~I.~Zwiebel, \textit{Charging the Superconformal Index}, {\tt arXiv:1111.1773 [hep-th]}.

\bibitem{Benini:2011nc}
F.~Benini, T.~Nishioka and M.~Yamazaki,
\textit{$4d$ Index to $3d$ Index and $2d$ TQFT}, {\tt arXiv:1109.0283 [hep-th]}.

\bibitem{Nakayama:2011pa}
Y.~Nakayama, \textit{$4D$ and $2D$ superconformal index with surface operator},
JHEP {\bf 1108} (2011) 084.

\bibitem{Dimofte:2011py}
T.~Dimofte, D.~Gaiotto and S.~Gukov,
\textit{$3$-Manifolds and $3d$ Indices},
{\tt arXiv:1112.5179 [hep-th]}.

\bibitem{Gang:2012yr}
D.~Gang, E.~Koh and K.~Lee,
\textit{Line Operator Index on $S^1\times S^3$},
{\tt arXiv:1201.5539 [hep-th]}.

\bibitem{Rains}  E. M. Rains, \textit{Transformations of elliptic hypergeometric
integrals},  Ann. of Math. {\bf 171} (2010), 169--243.

\bibitem{Hooft} G. 't Hooft, \textit{Naturalness, chiral symmetry,
and spontaneous chiral symmetry breaking},
Recent Developments in Gauge Theories (Eds. G. 't Hooft et. al.),
Plenum Press, New York, 1980, pp. 135--157.

\bibitem{spi:cirm} V. P. Spiridonov, \textit{Elliptic hypergeometric terms},
SMF S\'eminaire et Congr\`es
{\bf 23} (2011), 385--405; {\tt arXiv:1003.4491 [math.CA]}.

\bibitem{DSV}
F.~A.~H.~Dolan, V.~P.~Spiridonov and G.~S.~Vartanov,
\textit{From $4d$ superconformal indices to $3d$ partition functions},
Phys. Lett. {\bf B704} (2011), 234--241.

\bibitem{FV} G. Felder and A. Varchenko,
\textit{The elliptic gamma function and $SL(3, \mathbb{Z})\ltimes\mathbb{Z}^3$},
Adv. Math. {\bf 156} (2000), 44--76.

\bibitem{unit} J. F. van Diejen and V. P. Spiridonov, \textit{Unit circle
elliptic beta integrals}, Ramanujan J. {\bf 10} (2005), 187--204;
{\tt math.CA/0309279}.

\bibitem{Brodie:1998vv}
J.~H.~Brodie, P.~L.~Cho and K.~A.~Intriligator,
\textit{Misleading anomaly matchings?},
Phys. Lett. {\bf B429} (1998), 319--326.

\bibitem{KS} D. Kutasov and A. Schwimmer,
\textit{On duality in supersymmetric Yang-Mills theory}, Phys. Lett.
{\bf B354} (1995), 315--321.

\bibitem{sudano}
M.~Sudano,
\textit{The Romelsberger Index, Berkooz Deconfinement, and
Infinite Families of Seiberg Duals}, {\tt arXiv:1112.2996 [hep-th]}.

\bibitem{HW1}
M.~Henningson and N.~Wyllard,
\textit{Low-energy spectrum of $\mathcal{N} = 4$ super-Yang-Mills
on $T^3$: flat connections, bound states at threshold, and $S$-duality},
JHEP {\bf 0706} (2007) 001.

\bibitem{RSF}
A. G. Reiman, M. A. Semenov-Tian-Shansky and L. D. Faddeev,
\textit{Quantum anomalies and cocycles on gauge groups},
Funkt. Analiz i ego Pril. {\bf 18} (4) (1984), 64--72
(Funct. Analysis and its Appl.  {\bf 18} (4) (1984), 319--326).

\bibitem{Harvey}
J.~A.~Harvey, \textit{TASI 2003 lectures on anomalies}, {\tt hep-th/0509097}.

\end{thebibliography}
\end{document}